\documentclass[twocolumn]{aastex631}

\usepackage{orcidlink} 
\usepackage[T1]{fontenc}
\usepackage{ae,aecompl}

\usepackage{graphicx}	
\usepackage{amsmath}	
\usepackage{amssymb}	
\usepackage{float}
\usepackage{natbib}

\usepackage{hyperref}
\hypersetup{
	colorlinks=true,
	urlcolor=blue,    
	linkcolor=black,  
	citecolor=blue,   
}

\defcitealias{Pittordis_2018}{PS18}
\defcitealias{Pittordis_2019}{PS19}
\defcitealias{Badry_2021}{ERH}
\defcitealias{Banik_2018_Centauri}{BZ18}

\hypersetup{citecolor=blue, 
            linkcolor=red, 
            menucolor=blue, 
            urlcolor=blue}  

\newcommand{\cm}{{~\rm cm}}
\newcommand{\m}{{~\rm m}}

\newcommand{\s}{{~\rm s}}
\newcommand{\h}{{~\rm h}}

\newcommand{\g}{{~\rm g}}
\newcommand{\G}{{~\rm G}}
\newcommand{\K}{{~\rm K}}

\newcommand{\yr}{{~\rm yr}}

\newcommand{\zams}{\mathrm{ZAMS}}

\begin{document}

\title{Enabling high mass accretion rates onto massive main sequence stars by outer envelope mass removal}


\author{Ariel Scolnic \orcidlink{https://orcid.org/0009-0008-7967-2139}} \author{Ealeal Bear}\author{Noam Soker\,\orcidlink{0000-0003-0375-8987}}
\affiliation{Department of Physics, Technion, Haifa, 3200003, Israel; arielscolnic@campus.technion.ac.il; ealeal44@technion.ac.il; soker@physics.technion.ac.il}

\begin{abstract}
Using the one-dimensional numerical code \textsc{mesa}, we simulate mass accretion at very high rates onto massive main sequence stars, $M_{\rm ZAMS} =30,~60,~80 M_\odot$, and find that these stars can accrete up to $\simeq 10 \%$ of their mass without expanding much if we consider a simultaneous mass removal by jets. In this jetted-mass-removal accretion scenario, the accretion is through an accretion disk that launches jets. When the star expands due to rapid mass accretion, it engulfs the inner zones of the accretion disk and the jets it launches. We assume that these jets remove the outer layers of the envelope. We mimic this in the one-dimensional numerical code by alternating mass addition and mass removal parts. We add mass and energy, the accretion energy, to the outer layers of the envelope, leading to rapid stellar expansion. When the star expands by a few tens of percent, we stop mass addition and start mass removal until the star returns to its initial radius. We also show that the density of the accretion disk is larger than the density of the outer layers of the inflated envelope, allowing the disk to launch jets inside the outer inflated envelope layers. Our results show that main sequence stars can accrete mass at high rates while maintaining the deep potential well, as some models of eruptive systems require, e.g., some luminous red novae, the grazing envelope evolution, and the 1837-1856 Great Eruption of Eta Carinae. 
\end{abstract}


\section{Introduction}
\label{sec:intro}

Intermediate Luminosity Optical Transients (ILOTs) are transient objects powered by mass accretion onto a main sequence or slightly evolved star. The accreted mass may come from a companion star via mass transfer or a merger process, both occurring at high rates. ILOTs typically have peak luminosities between those of classical novae, or even slightly below classical novae, and those of supernovae (e.g., \citealt{Mouldetal1990, Bondetal2003, Rau2007, Ofek2008, Masonetal2010, Kasliwal2011, Tylendaetal2013, Kasliwaletal2012, Kaminskietal2018, BoianGroh2019, Caietal2019, Jencsonetal2019, Kashietal2019Galax, PastorelloMasonetal2019, Blagorodnovaetal2020, Banerjeeetal2020, Howittetal2020, Jones2020, Kaminskietal2020Nova1670, Kaminskietal2021Nova1670, Klenckietal2021, Stritzingeretal2020AT2014ej, Stritzingeretal2020SNhunt120, Blagorodnovaetal2021, Mobeenetal2021, Pastorelloetal2021, Pastorelloetal2023, Addisonetal2022, Caietal2022,  Wadhwaetal2022, Kaminskietal2023, Karambelkaretal2023,  ZainMobeenetal2024, Kaminski2024}).
\textit{We use the term ILOTs for all transients powered by gravitational energy,} (for earlier use of this term see, e.g., \citealt{Berger2009, KashiSoker2016Terms, MuthukrishnaetalM2019}). Other terms with partial (but not complete) overlap with ILOTs include luminous red novae, red novae, intermediate luminosity red transients, or gap transients (e.g., \citealt{Jencsonetal2019, Caietal2022b}). There is also a diversity in classifying the subclasses of these transients (e.g., \citealt{KashiSoker2016Terms, PastorelloMasonetal2019, PastorelloFraser2019}). 
We consider luminous red novae to be ILOTs powered by a complete merger that leaves one stellar remnant \citep{KashiSoker2016Terms}. {{{{ Considering the emission processes, the common envelope evolution (CEE) might also be considered as a complete merger because there is only one photosphere, that of the common envelope. Later, the common envelope is ejected, and the core and companion emerge so from the dynamical point of view it is not a complete merger. }}}}

High mass transfer rates from a giant star onto a main sequence companion can occur during grazing envelope evolution (GEE) and CEE. In the case of eccentric orbits, the jet-launching might be along the entire orbit or near periastron passages. Although the accreting companion might, in principle, be sub-stellar, i.e., a planet or a brown dwarf (e.g., \citealt{RetterMarom2003,  Metzgeretal2012, Yamazakietal2017, Kashietal2019Galax, Gurevichetal2022, Deetal2023, Oconnoretal2023}), we here consider main sequence companions, as most studies do (e.g., \citealt{Tylendaetal2011, Ivanovaetal2013a, Nandezetal2014, Kaminskietal2015, Pejchaetal2016a, Pejchaetal2016b, Soker2016GEE, Blagorodnovaetal2017, MacLeodetal2017, MacLeodetal2018,  Segevetal2019, Howittetal2020, MacLeodLoeb2020, Qianetal2020, Schrderetal2020, Blagorodnovaetal2021, Addisonetal2022, Zhuetal2023, Tylendaetal2024}). Energy sources during the CEE might be a recombination of the ejected gas (e.g., \citealt{MatsumotoMetzger2022}), dynamical interaction of the companion with an envelope of the giant star (which causes spiraling-in of the companion), and mass accretion onto the companion. 

Recombination of the ejected gas and the collision of ejected envelope gas with earlier ejected gas in and near the equatorial plane heat the gas and power radiation (e.g., \citealt{Pejchaetal2016a, Pejchaetal2016b, Pejchaetal2017, MetzgerPejcha2017, HubovaPejcha2019}). We, on the other hand, consider mass accretion onto the compact companion via an accretion disk and the launching of jets by this disk to be the primary energy source of the radiation of bright ILOTs (e.g., \citealt{Soker2020ILOTjets, SokerKaplan2021RAA}). Observed ILOTs with bipolar morphologies (e.g., \citealt{Kaminski2024}) support the notion that jets power most, or even all, bright ILOTs (e.g.,  \citealt{Soker2023BrightILOT, Soker2024}). Systems with neutron stars or black holes as the mass-accreting companion might appear as core-collapse supernovae (e.g., \citealt{SokerGilkis2018, Gilkisetal2019, GrichenerSoker2019, YalinewichMatzner2019, Schreieretal2021}; for review see \citealt{Grichener2025}).

The bipolar morphology of the material that the Luminous blues variable Eta Carinae ejected in the great Eruption of the nineteenth century, the Homunculus, is compatible with a jet-powering of the Great Eruption  (e.g., \citealt{Soker2001, KashiSoker2010}). The jet-powering scenario requires that the companion's (secondary) mass of Eta Carinae is $M_2 \simeq 30- 80 M_\odot$ \citep{KashiSoker2016EtaCar}, and that during the twenty-year-long Great Eruption, the companion accreted a mass of $\simeq 4 M_\odot$ \citep{KashiSoker2010}. To power an energetic event, the companion must accrete large amounts of mass without expanding much.  
 In a previous exploratory study, we \citep{BearSoker2025} showed that massive main-sequence stars can avoid expansion while accreting mass at high rates if there is a mechanism to eject the outer parts of accreted mass simultaneously with mass accretion. We considered the mass removal mechanism to be jets that the star launches. The present study uses different criteria for a more realistic scenario to achieve this mass removal (Section \ref{sec:Method}). Our results strengthen the earlier finding of the mechanism to prevent stellar expansion (Section \ref{sec:Results}). In Section \ref{sec:Disk}, we calculate the properties of the accretion disk, which shows it can exist in the outer zones of the envelope. In Section \ref{sec:EjectedMass} we discuss some properties of the ejected mass. We summarize in Section \ref{sec:Summary}.

\section{Method}
\label{sec:Method}

We used version 24.03.1 of the stellar evolution code Modules for Experiments in Stellar Astrophysics (\textsc{mesa}; \citealt{Paxtonetal2011, Paxtonetal2013, Paxtonetal2015, Paxtonetal2018, Paxtonetal2019, Jermynetal2023}) in its single star mode. We base our simulations on the example of \textit{zams\_to\_cc\_80}. 
We use this example until we start accreting mass; all other parameters remain as in the default of \textsc{mesa}.
 
We evolve three stellar models with an initial zero-age main sequence (ZAMS) mass of $M_{\rm ZAMS}=80M_\odot$, $M_{\rm ZAMS}=60M_\odot$ and $M_{\rm ZAMS}=30M_\odot$, and metallicity of $z=0.0142$. 
We start mass accretion and energy deposition on the main sequence at the age of $t_{\rm MS}= 2.85 \times 10^5 \yr$, $t_{\rm MS}= 3.50 \times 10^5 \yr$ and $t_{\rm MS}= 6.50 \times 10^5 \yr$ for $M_{\rm ZAMS}=80M_\odot$, $M_{\rm ZAMS}=60M_\odot$ and  $M_{\rm ZAMS}=30M_\odot$, respectively. When we start the accretion process, we set the time $t_{\rm acc}=0$. 

We accrete mass in pulses, each comprising two parts. In the first part, we add mass at a rate of $\dot{M}_{\rm add}=1.5 \times 10^{-2} M_\odot \yr^{-1}$, and simultaneously, we deposit energy at a fraction $\eta_{\rm acc}$ of the gravitational energy that the added mass releases.
{{{{ Specifically, we deposit energy during the mass addition part (but not the mass removal part) at a rate of $\dot E_{\rm add} = \eta_{\rm acc} \dot M_{\rm add} [G M(t)/R(t)]$, where $M(t)$ and $R(t)$ are the mass and radius of the star, respectively. We deposit this energy in the envelope's outer $0.1$ by radius, with a uniform power per unit mass in this zone. }}}} When the star expands to a radius of 
\begin{equation}
    R_{\rm add}=k_RR_0 , 
  \label{eq:Radd}
\end{equation} 
we start the second part of the pulse of mass removal; $R_0$ is the initial radius of the star at $t_{\mathrm{MS}}$, and in this study, we calculate for either $k_R=1.17$ or $k_R=1.3$. We remove mass at a rate of $\dot{M}_{\rm add}=-1 \times 10^{-2} M_\odot \yr^{-1}$. 
Once the star contracts back to the initial radius $R_0$, we stop mass removal and start mass and energy addition of the next pulse. 
Numerically, we set the initial time step at every pulse to start at 
$\textit{years\_for\_initial\_dt} = 5 \times 10^{-7} \, \yr$ for both the removal and accretion parts of the pulse, for 
$M_{\mathrm{ZAMS}} = 80M_\odot$, $M_{\mathrm{ZAMS}} = 60M_\odot$, and $M_{\mathrm{ZAMS}} = 30M_\odot$.
In this study, we keep the maximum
 (equation \ref{eq:Radd}), and minimum, $R_0$, stellar radii constant, as compared with our previous study \citep{BearSoker2025}, where the added and removed mass were constant. 
We think it is more realistic because the interaction with the companion might truncate the radius of the mass-accreting star. In reality, the star radius stays at $\simeq k_R R_0$, and the inflow and outflow occur simultaneously. 

{{{{ With the specific scheme we apply the \textsc{mesa} stellar code cannot handle much higher mass addition rates than $\dot{M}_{\rm add}=1.5 \times 10^{-2} M_\odot \yr^{-1}$. Future one-dimensional studies will examine different mass addition and removal schemes to allow higher mass addition rates. Eventually, three-dimensional simulations should examine the maximum mass accretion rate possible.   }}}}

\section{Results}
\label{sec:Results}

\begin{figure}[]
\centering
\includegraphics[trim=0cm 1cm 0cm 0cm, clip, width=\linewidth]{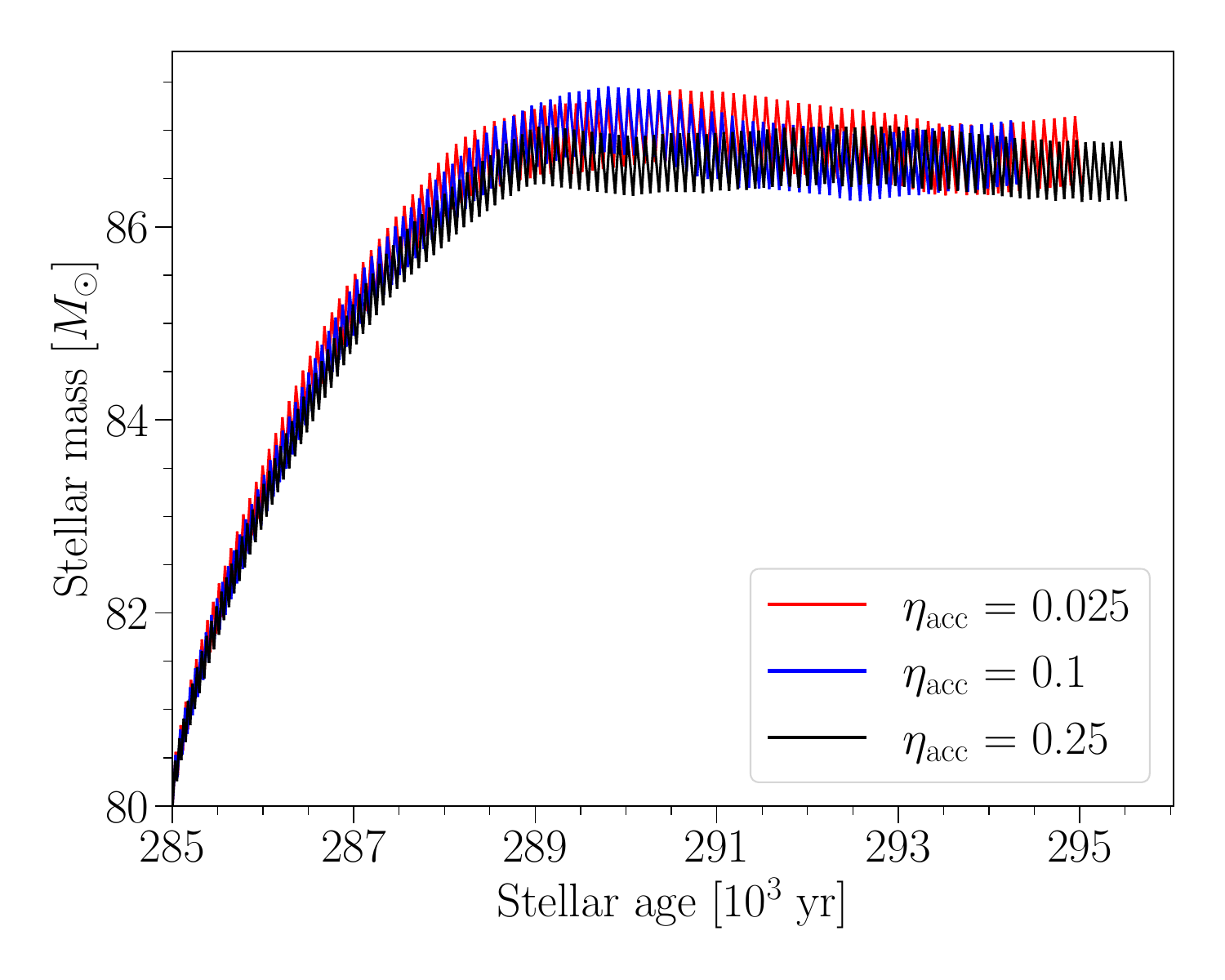}
\caption{Stellar mass as a function of stellar age for the $M_{\rm{ZAMS}}$ = 80 $M_\odot$ model. The graphs start when we start the accretion process. Red, blue, and black lines signify different energy deposition efficiencies of $\eta_{\rm acc}=0.025, 0.1,0.25$, the fraction of the gravitational energy of the added mass. All cases are for $R_{\rm add} = 1.3 R_0$ ($k_R=1.3$ in equation \ref{eq:Radd}).  For the scheme we use here, the energy deposition efficiency has little effect on the outcomes.}
\label{fig:MassTime80E}
\end{figure}

In Figure \ref{fig:MassTime80E}, we present the stellar mass as a function of stellar age for the $M_{\rm ZAMS}=80M_\odot$ stellar model and for three values of $\eta_{\rm acc}$, which is the fraction of the gravitational energy that the added mass releases to the envelope. In these simulations, the radius where we turn from mass addition to mass removal is $k_R=1.3$ in equation (\ref{eq:Radd}). We learn the following from these simulations. (1) The added energy does not influence the results much, at least in the scheme by which we accrete mass. (2) The graph shows that the mass fluctuates due to our two-part mass accretion scheme, one of mass addition and one of mass removal. (3) The net mass accretion rate is very high at early times and decreases to zero. (4) Most relevant to our goals, the mass reaches an asymptotic value significantly above the initial mass.     

\begin{figure}[]
	\centering
\includegraphics[trim=0cm 0.5cm 0cm 0.5cm ,clip, width=\linewidth]{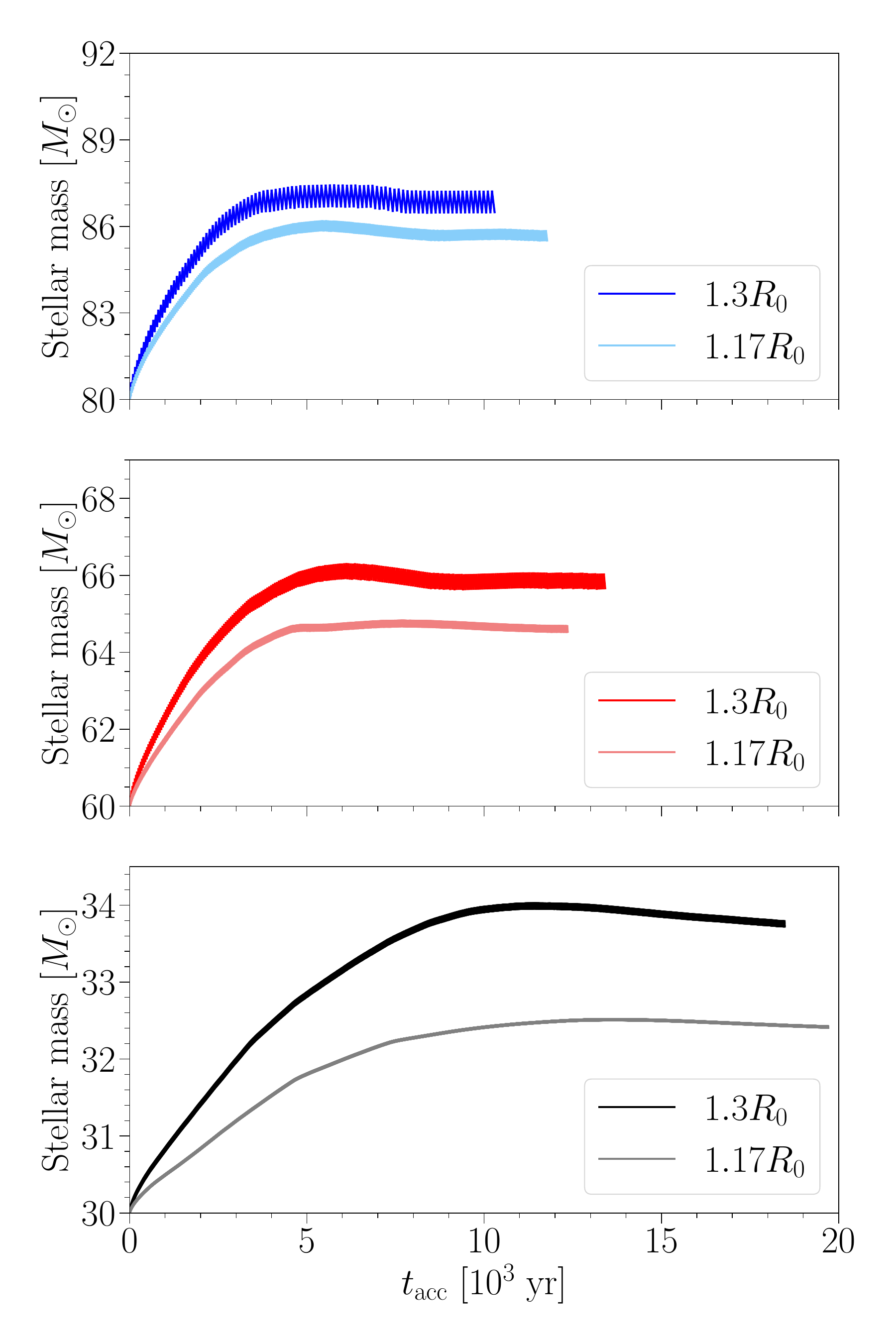}
\caption{Stellar mass as a function of time, measured from the beginning of the accretion process, for three stellar models with $M_{\rm{ZAMS}}= 80M_\odot$,$M_{\rm{ZAMS}}= 60M_\odot$, and $M_{\rm{ZAMS}}= 30M_\odot$, and for two values of the stellar radius where we change from mass addition to mass removal, $R_{\rm add} = 1.17 R_0$ and $R_{\rm add} = 1.3 R_0$.
The age of the stellar models on the main sequence when pulses of mass addition and removal start are $t_{\rm{MS}} (80 M_\odot) = 2.85\times 10^5$, $t_{\rm{MS}}(60 M_\odot)= 3.50 \times 10^5$ and $t_{\rm{MS}} (30 M_\odot) = 6.50 \times 10^5$ $\yr$. 
All simulations are for mass addition rate of $\dot{M}_{\rm{add}}=1.5 \times 10^{-2} {M_\odot}\yr^{-1}$ and mass removal rate of $\dot{M}_{\rm{rem}}=-1 \times 10^{-2} {M_\odot}\yr^{-1}$, and with energy injection efficiency of $\eta_{\rm acc}=0.1$. 
Note that the thickness of each plot is not the width of the line but is a result of the very dense spacing of the mass addition-removal pulses (see Figure \ref{fig:MassTime80E}); namely, the upper boundary of the line is the end of the mass addition parts of the pulses, and the lower boundary is the end of the mass removal parts.  } 
\label{fig:MassTimeAll}
\end{figure}

In Figure \ref{fig:MassTimeAll}, we present the stellar mass as a function of time for the three stellar models and two maximum radii, namely, $k_R=1.17$ and $k_R=1.3$ in equation (\ref{eq:Radd}). In these simulations $\eta_{\rm acc} =0.1$. 
The three models in the six simulations behave similarly, and the asymptotic value of the final mass is larger for a larger maximum radius. 
In Table \ref{tab:Step1_table}, we summarize the relevant properties of the six simulations that we present in Figure \ref{fig:MassTimeAll}. We present the average mass accretion properties when the stars increase their mass by $5\%$. This value is well before the asymptotic mass, and it is further motivated by the model where during the 20-years-long Great Eruption of Eta Carinae, a companion of $\simeq 80 M_\odot$ accreted $\simeq 4 M_\odot$ \citep{KashiSoker2016EtaCar}.
\begin{table*}   
\begin{center}
\caption{Properties of mass-accreting stellar models}
\label{tab:Step1_table}
\renewcommand{\arraystretch}{1.5} 
\begin{tabular}{|c|c|c|c|c|c|c|}
\hline
 \textbf{$M_{\rm{ZAMS}}$}& \multicolumn{2}{|c|}{\textbf{$80 M_\odot$}} & \multicolumn{2}{|c|}{\textbf{$60 M_\odot$}} & \multicolumn{2}{|c|}{\textbf{$30 M_\odot$}} \\ \hline
 \textbf{$R_{\rm{add}}$}& \textbf{$1.3 R_0$} & \textbf{$1.17 R_0$} & \textbf{$1.3 R_0$} & \textbf{$1.17 R_0$} & \textbf{$1.3 R_0$} & \textbf{$1.17 R_0$} \\ \hline
\textbf{$M_{\rm{acc,f}} [M_\odot]$} & 7.5 & 5.9 & 6.3 & 4.9 & 4.0 & 2.6 \\ \hline
\textbf{$M_{\rm{add,5}} [M_\odot]$}& 10.0& 13.2& 9.9& 13.9& 13.6& 24.3\\ \hline
\textbf{$M_{\rm{rem,5}} [M_\odot]$}& 5.9& 9.1& 6.8& 10.8& 12.1& 22.8\\ \hline 
 \textbf{$M_{\rm{acc,5}} [M_\odot]$}& 4.1& 4.1& 3.1& 3.1& 1.5&1.5\\ \hline
\textbf{$t_{\rm{add,5}} [\rm{yr}]$}& 666& 881& 659& 925& 908& 1623\\ \hline
\textbf{$t_{\rm{rem,5}} [\rm{yr}]$}& 586& 912& 681& 1081& 1210& 2284\\ \hline
\textbf{$t_{5} [\rm{yr}]$} & 1252& 1793& 1340 & 2006 & 2118 & 3907 \\ \hline
\textbf{$\dot{M}_{\rm{acc,5}} [M_{\odot} \yr^{-1}]$} & $33.1\times 10^{-4}$ & $22.8\times 10^{-4}$ & $22.9\times 10^{-4}$ & $15.3\times 10^{-4}$ & $7.2\times 10^{-4}$ & $3.9\times 10^{-4}$ \\ \hline
\textbf{$N_{\rm{P,5}}$} & 18 & 61 & 36 & 113 & 359 & 2583 \\ \hline
\end{tabular}
\end{center}
Notes: Table summarizing six simulations. The first two rows are the input parameters, the initial stellar mass $M_{\zams}$, and the maximum radius (equation \ref{eq:Radd}). In these six simulations $\eta_{\rm acc}=0.1$.
$M_{\rm{acc,f}}$ is the net asymptotic accreted mass (the maximum mass the star can gain in the numerical scheme we use).
The rest are results when the star reaches a mass increase of $5\%$. 
$M_{\rm{add,5}}$ and $M_{\rm{rem,5}}$ are the total added mass and removed mass when the stellar mass reaches an increase by $M_{\rm acc,5}=5\%$ (up to the accuracy of a pulse), i.e., $M_{\rm acc,5}=0.05M_{\rm ZAMS} =M_{\rm{add,5}} - M_{\rm{rem,5}}$. 
$t_{\rm{add,5}}$ and $t_{\rm{rem,5}}$ are the total time in the mass-addition and mass-removal parts, respectively, $t_5= t_{\rm{add,5}}+t_{\rm{rem,5}}$, and $\dot M_{\rm acc,5} = 0.05M_{\rm ZAMS}/t_5$ is the average mass accretion rate until the star mass is $1.05M_{\rm ZAMS}$. $N_{\rm{P,5}}$ is the total number of pulses until that time. 
\end{table*}
 
\begin{figure}[]
	\centering
\includegraphics[trim=0cm 0cm 0cm 0cm ,clip, width=\linewidth]{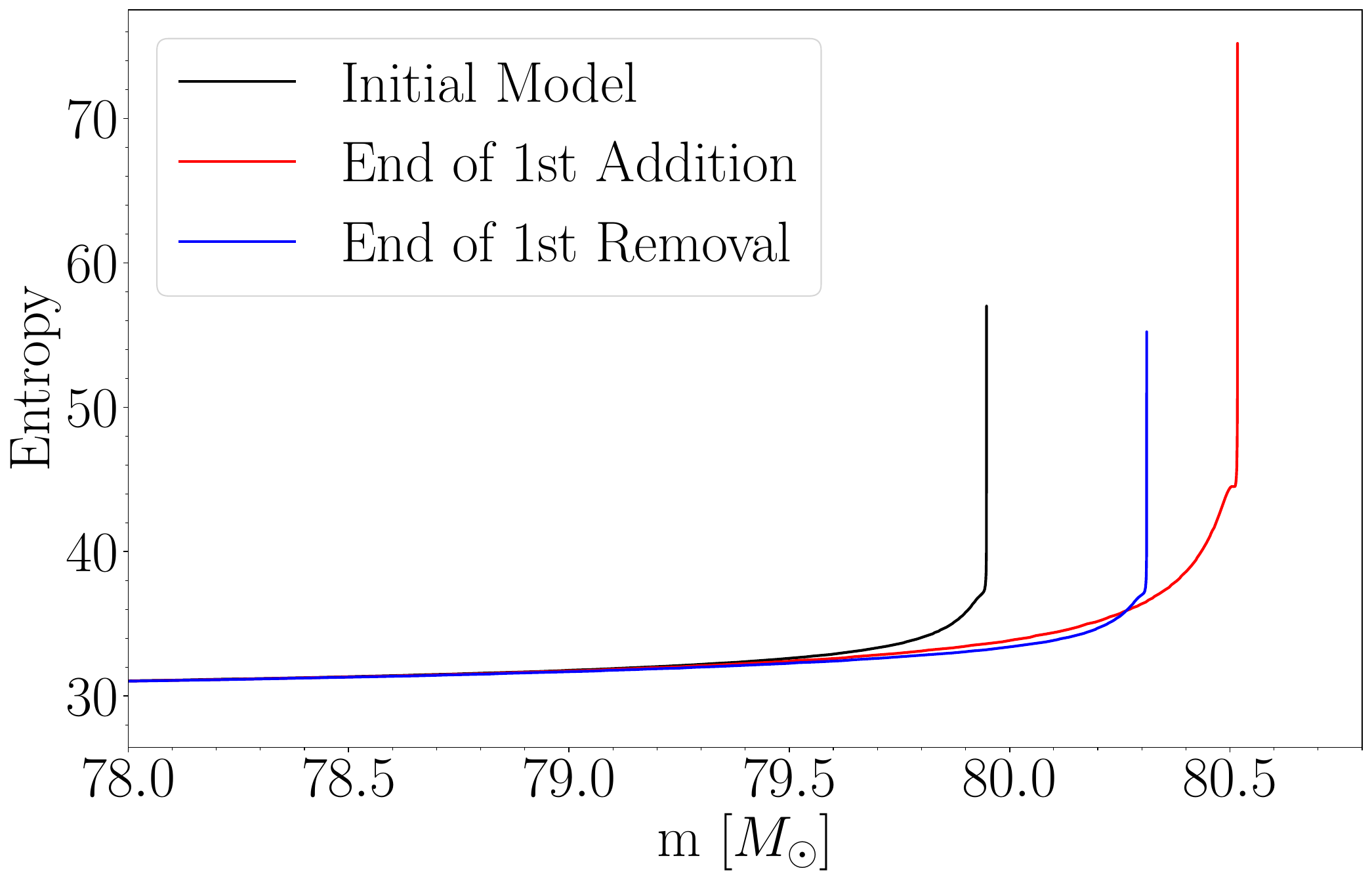}
\caption{ {{{{ Entropy profiles as a function of mass at three stages of the first accretion pulse of the $M_{\rm ZAMS} = 80 M_\odot$ model with $R_{\rm add} = 1.3R_0$ and $\eta_{\rm acc}=0.1$.  
The black, red, and blue lines represent the entropy profiles of the model at $t_{\rm acc}=0$ (just before accretion starts), end of mass addition phase of the first pulse, and the end of the mass removal phase of the first pulse, respectively.
 Entropy is given by \textsc{mesa} as the specific entropy divided by  $N_{\rm A} k_{\rm B}$, where $N_{\rm A}$ is Avogadro's number and $k_{\rm B}$ is the Boltzmann constant. }}}}
 }
\label{fig:Entropy}
\end{figure}

The main conclusion of our simulations is that mass removal from the stellar outskirts allows massive stars to accrete up to $\simeq 10\%$ of their mass when their radius is truncated at $k_R = 1.3$ times their initial radius, without an increase in their final stellar radius. The maximum mass is reached asymptotically. Lower stellar mass can grow by a larger factor than more massive stars. Larger maximum radii allow the star to accrete mass at a higher rate and reach higher asymptotic mass. We keep the maximum stellar radius small to maintain a deep potential well so that the accretion disk launches energetic jets. As \cite{BearSoker2025} showed, the outer parts have very high entropy, such that their removal causes the envelope to shrink. {{{{ We demonstrate the high entropy of the accreted mass in figure \ref{fig:Entropy}. We present the entropy profiles as a function of mass at three times of the first pulse of the $\eta_{\rm acc}=0.1$ case that we present in Figure \ref{fig:MassTime80E}: before mass addition, at the end of mass addition, and at the end of mass removal. The outer zone of the envelope after mass addition has high entropy (red line). We remove this mass in the mass removal part. }}}} 
\cite{BearSoker2025} added and removed mass in pre-determined steps of mass addition and removal, while here, we set maximum and minimum stellar radii, which we consider more realistic. When the high-entropy outer envelope regions are not removed, the star expands unstably for the accretion rates we simulate here, as \cite{SchurmannLanger2024} obtained in their extensive study of main sequence stars accreting at constant rates without mass removal. 

\cite{BearSoker2025} accrete small amounts of mass, $<1\%$ of the stellar mass,  at accretion rates of $0.01 M_\odot \yr^{-1}$ and $0.0025 M_\odot \yr^{-1}$, for the $M_{\rm ZAMS} = 60 M_\odot$ and $M_{\rm ZAMS} = 30 M_\odot$ stellar models, respectively. These accretion rates are higher than the average ones we obtain until the star increases its mass by $5\%$. As figures \ref{fig:MassTime80E} and \ref{fig:MassTimeAll} show, the mass accretion rate decreases with time; at early times, the accretion rates we find here are similar to those that \cite{BearSoker2025} obtained. 

When we add mass continuously but add no energy, the star slowly expands. For example, adding mass without energy in the $M_{\rm ZAMS}=30 M_\odot$ at a rate of $10^{-3} M_\odot \yr^{-1}$ causes the star to expand by $9.4 \%$ after accreting $3.0 M_\odot$. However, we think mass accretion without energy accretion is not realistic as the added mass must release energy when accreted. 

\section{The accretion disk}
\label{sec:Disk}
In the scenario we propose, the accretion disk around the star launches jets (or disk wind) that remove the outer layer of the inflating envelope. This requires that as the envelope expands, it does not destroy the disk immediately; i.e., the accretion disk continues to exist inside the outer zones of the inflated envelope. This allows the disk to launch jets in the outer zone of the envelope; the jets remove high-entropy envelope mass from this region. The condition of a disk inside the star is that the density in the disk is higher than that in the outer envelope. 

We use the scaled equation for the density in the disk from  \cite{FrankKingRaine2002}, their equation 5.49,  
and scale it to our typical parameters. This yields the equation for the typical density of the accretion disk near the stellar surface 
\begin{equation}
\begin{split}
\rho_{\rm d} & =  8.4 \times 10^{-7} 
\left( \frac{ \alpha_{\rm d}}{0.1} \right) ^{-7/10}
\left( \frac {\dot M_{\rm acc}}{10^{-3} M_\odot \yr^{-1}}    \right)^{11/20} 
\\ & \times
\left( \frac {M}{60 M_\odot} \right)^{5/8} 
\left( \frac {r}{10 R_\odot} \right)^{-15/8} 
\left( \frac {f}{0.5} \right) ^{11/5} \g \cm^{-3} ,
\label{eq:RhoDisk}
\end{split}
\end{equation}
where $\alpha_{\rm d}$ is the Shakura-Sunyaev viscosity parameter, $\dot{M}_{\rm acc}$ is the mass accretion rate through the disk, $M$ is the stellar mass, $r$ is the radius in the disk, and $f$ is given by 
\begin{equation}
f = \left(1 - \sqrt{\frac{R_0}{r}}\right)^{\frac{1}{4}}  ,
\label{eq:f}
\end{equation}
where $R_0$ is the stellar radius before it expands following the rapid mass accretion. 
In equation (\ref{eq:RhoDisk}), we scaled the factor $f$ at a disk radius of $r=1.14R_0$, which is smaller than the expansion of the star following the mass addition, i.e., this part of the disk that is inside the inflated envelope. 
Equation (\ref{eq:RhoDisk}) assumes a constant mass accretion rate through the disk and an isolated disk. In the scenario we study, the disk launches jets and interacts with the inflated stellar envelope. We should keep in mind these uncertainties. 

\begin{figure}[]
\centering
\includegraphics[trim=0cm 0cm 0cm 0cm ,clip, width=\linewidth]{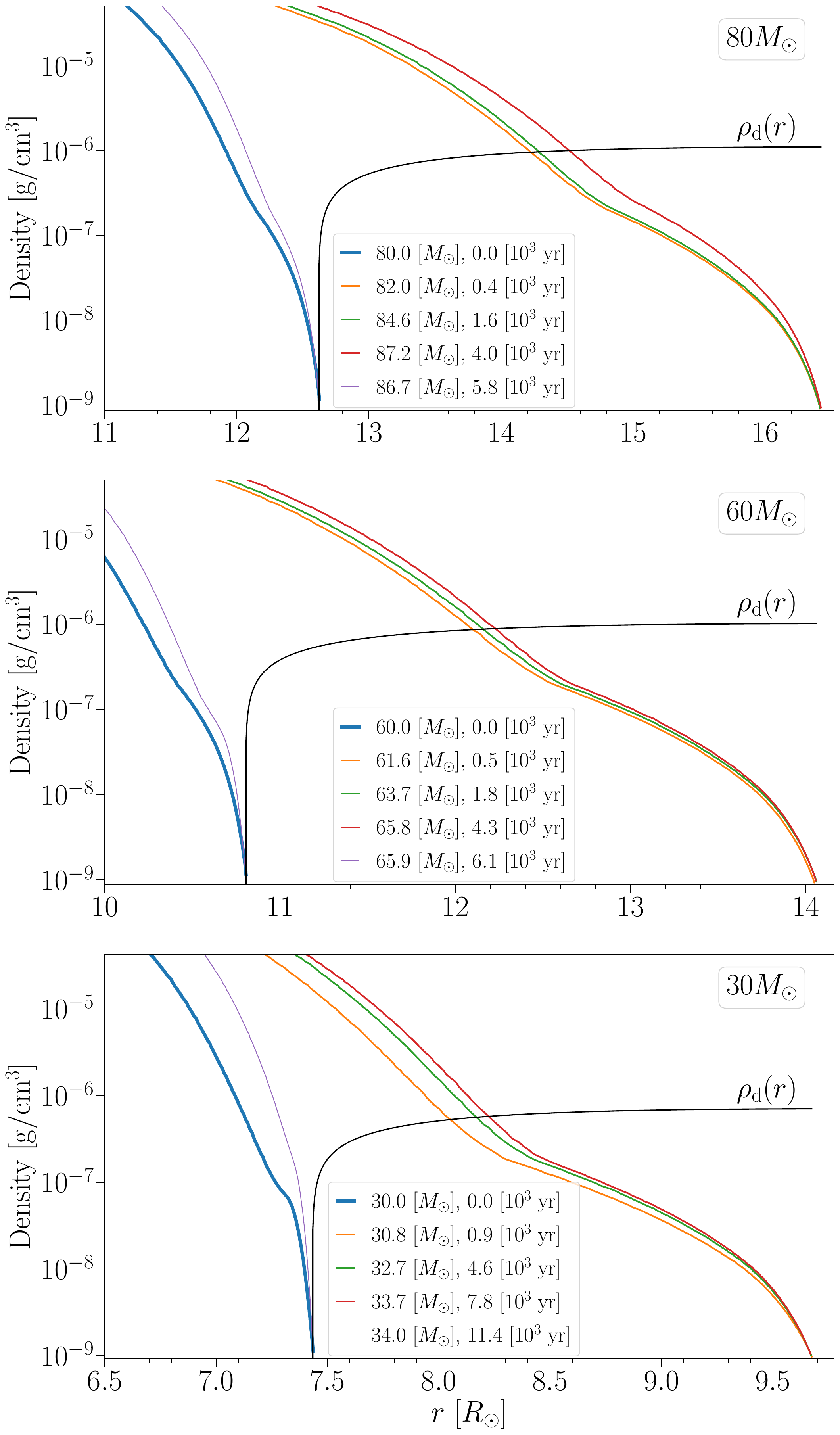}
\caption{Density profiles at five different times for the three stellar models. The insets show the stellar mass as well as the time from the start of the accretion process for each line.  
The blue line represents the density profile at $t_{\rm acc}=0$. 
The red, green, and orange lines represent the density profiles at the end of the mass addition part of three pulses. The purple line represents the density profile after the mass removal part of the last pulse in the simulation. The black line is the accretion disk density according to equation (\ref{eq:RhoDisk}) with $\alpha_{\rm d}=0.1$; for each model, we take the initial stellar mass and the average accretion rate until the star mass increases by $5 \%$.  All cases are for $\eta_{\rm acc}=0.1$ and  $R_{\rm add} = 1.3 R_0$.
}
\label{fig:Density}
\end{figure}

In Figure \ref{fig:Density}, we present density profiles for the three stellar models, at five different times during the accretion process. We also present the disk density according to (\ref{eq:RhoDisk}). 
We take the stellar mass to be the initial mass and the accretion rate to be the average one until the star reaches a mass of $1.05 M_{\rm ZAMS}$, i.e., $\dot M_{\rm acc,5}$, and the parameter $f$ according to equation (\ref{eq:f}). 

Figure \ref{fig:Density} shows that the density in the accretion disk is larger by an order of magnitude and more than the envelope density in the outer $\simeq 10\%$ of the envelope by radius. 
The higher disk density implies that the disk can robustly exist in the outer envelope; in our proposed scenario, the accretion disk launches jets that remove the outer envelope. We term this scenario 
jetted-mass-removal accretion. 

\section{The ejected mass}
\label{sec:EjectedMass}

\begin{figure}[]
	\centering
\includegraphics[trim=1.2cm 1cm 0cm 0cm ,clip, scale=0.3]{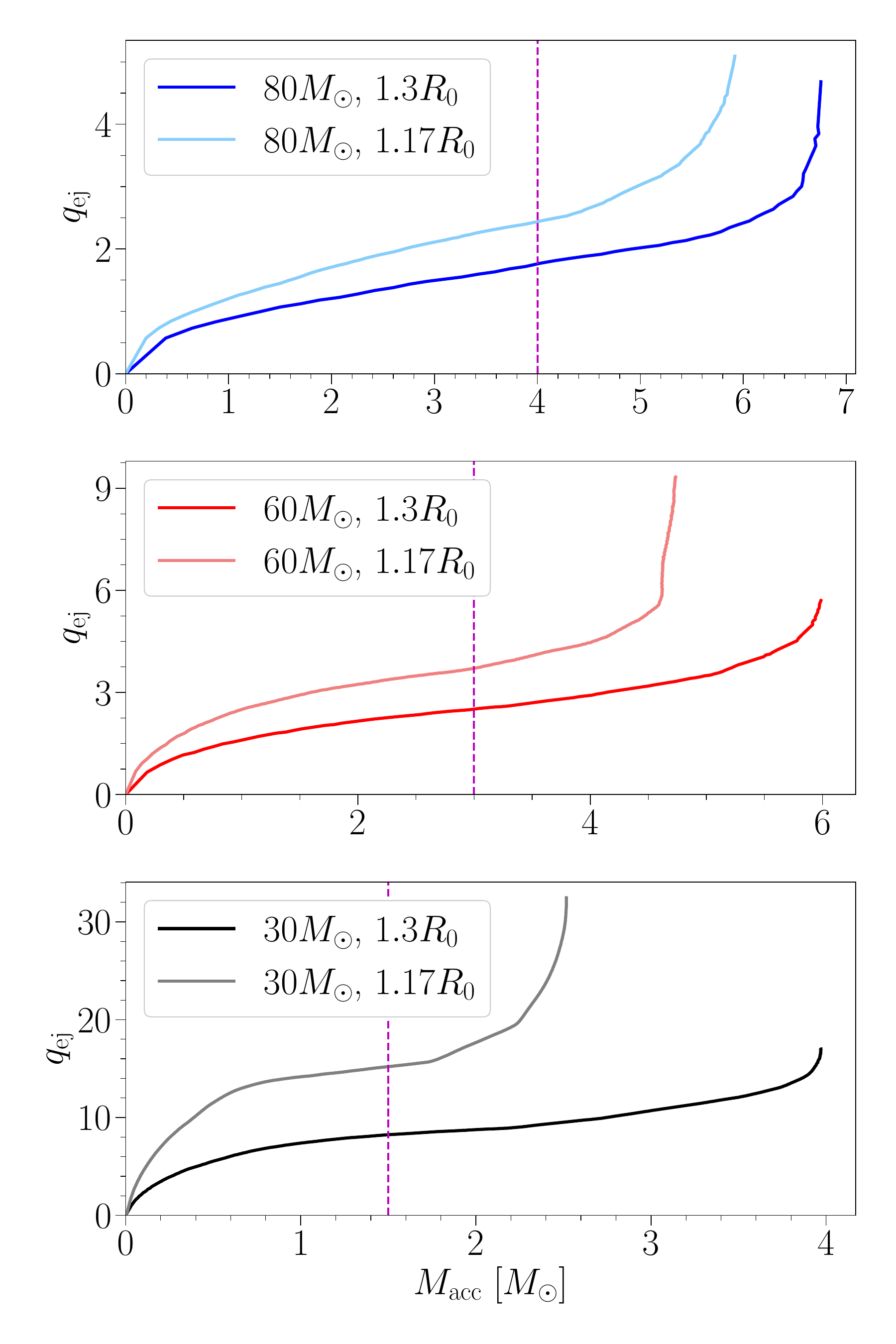}
\caption{The ratio of the removed mass to the net accreted mass (equation \ref{eq:MassRatio}) as a function of accreted mass for the six simulations we present in Figure \ref{fig:MassTimeAll}. The vertical dashed line marks the $5 \%$ mass growth. The insets show the initial stellar mass and the maximum radius (equation \ref{eq:Radd}).}
\label{fig:MassRatio}
\end{figure}

The jetted-mass-removal accretion scenario, where the mass-accreting non-degenerate star does not expand much, requires the mass that the jets remove from the envelope to be very large, up to several times the net accreted mass for the $M_{\rm ZAMS} = 30 M_\odot$ model. In Figure \ref{fig:MassRatio}, we plot the ratio of the ejected mass to the net accreted mass, 
\begin{equation}
 q_{\rm ej} (t) \equiv M_{\rm rem}(t)/M_{\rm acc} (t). 
    \label{eq:MassRatio}
\end{equation}
The ejected mass contains the removed mass and the mass in the jets. However, the simple one-dimensional simulations do not separate the jets' material from the envelope. 

We assume the accreted gas comes from a distance much larger than the stellar radius. {{{{ The removed mass originates from large distances and is carried away by the jets back to similarly large distances; therefore, it adds no net energy to the outflow. Only the accreted mass, $M_{\rm acc}=M_{\rm add} - M_{\rm rem}$ contributed energy to the removed mass. }}}}
With the virial theorem, the net energy that the accreted mass releases is $E_{\rm acc}=M_{\rm acc} v^2_{\rm es} /4$, where $v_{\rm es}$ is the escape velocity from the stellar surface.
{{{{ We assume that most of it, a fraction $\zeta$, goes to expel the removed mass, and only a small fraction $1-\zeta \ll 1$ goes to radiation. The last inequality is justified because the jet-envelope interaction occurs in a very optically thick gas (inside the photosphere). }}}} The above discussion implies that the average, by energy, {{{{ terminal (i.e., at large distances) }}}} velocity of the ejected material is 
\begin{equation}
\bar{v}_{\rm ej} =  \left( \frac{\zeta}{2 q_{\rm ej}} \right)^{1/2} v_{\rm es}.  
    \label{eq:Vej}
\end{equation}
The conclusion from the values of $q_{\rm ej}$ that we find is that the typical expansion velocity of the ejecta, when large amounts of mass are ejected, is significantly lower than the escape velocity from the star. The outflow velocity distribution is a subject of future three-dimensional hydrodynamical simulations. 


\section{Discussion and Summary}
\label{sec:Summary}
This study is the second study that explores the jetted-mass-removal accretion scenario. In this scenario, the accretion onto a non-degenerate star is through an accretion disk that launches jets. The star does not expand much even when the mass accretion rate is very high and includes energy accretion because the jets that the disk launches remove the high-entropy outer envelope regions. \cite{BearSoker2025} presented the principle of this process and its basic properties; they accreted $\lesssim 1 \%$ of the stellar mass. We used a different mass-accretion method and accreted more mass, $\simeq 10 \%$ of the stellar mass (Table \ref{tab:Step1_table}).  The accretion method has a part of mass addition at a constant rate until the star expands to a radius given by equation (\ref{eq:Radd}); at that time, the mass removal part starts until the star contracts back to its original radius. Without the mass removal part, the star expands by a large factor after accreting a small amount of mass. 

We found that our demand of no final radius expansion allows an accretion of $\simeq 10\%$ of the stellar mass, reached asymptotically (Figures \ref{fig:MassTime80E} and \ref{fig:MassTimeAll}). The fraction of stellar mass growth increases with decreasing stellar mass and increasing maximum radius allowed during the accretion process (Table \ref{tab:Step1_table}).

We also showed (Figure \ref{fig:Density}) that the accretion disk's density is larger than the density in the outer regions of the inflated envelope, implying the jets can exist in the outer inflated envelope. The accretion disk might launch jets that remove outer envelope zones.   

In Figure \ref{fig:MassRatio}, we presented the ratio of ejected mass to net accreted mass. The jetted-mass-removal accretion involves the removal of a large fraction of the added mass when the star accretes large amounts of mass, i.e., $\gtrsim 5 \%$ of its mass. This implies that the ejecta average {{{{ terminal velocity (velocity at large distances relative to the stellar radius) }}}} is much lower than the escape velocity from the star.

Although our study puts the jetted-mass-removal accretion scenario on solid ground, there is much to do. The following steps in exploring the jetted-mass-removal accretion scenario should be to extend the present study to other types of stars, like low-mass main sequence stars and WR stars, and eventually to conduct three-dimensional hydrodynamical simulations of an accretion disk around an inflated star. 
The primary motivation for conducting these studies is the model, according to which a companion star that accretes mass at a high rate and launches jets might explain many transients, such as ILOTs, luminous red novae, luminous blue variable major eruptions, and some pre-supernova outbursts.
We expect that three-dimensional simulations that include dynamical effects will allow higher mass accretion rates, as the accretion model requires to explain some transient objects.


\section*{Acknowledgments}

We thank an anonymous referee for helpful comments. A grant from the Pazy Foundation supported this research.



\label{lastpage}

\end{document}